\documentclass[reprint,amsmath,amssymb,aip,showpacs,pra]{revtex4-1}
\usepackage{graphicx}% Include figure files
\usepackage{dcolumn}% Align table columns on decimal point
\usepackage{bm}% bold math
\usepackage[usenames,dvipsnames]{color}
\usepackage[normalem]{ulem} %to strike through text
\usepackage[utf8]{inputenc}
\usepackage[hidelinks]{hyperref}% add hypertext capabilities
\graphicspath{{./Files/PDFs/}}

\begin{document}

%=========================================================================================

\title{Measurement of temperature of a dusty plasma from configuration}

%\vspace{8mm}
%\large
\author {Rupak Mukherjee}
\email{rmukherj@pppl.gov; rupakm@princeton.edu}
\affiliation{Princeton Plasma Physics Laboratory, Princeton, New Jersey, 08540, USA}
\affiliation{Institute for Plasma Research, HBNI, Gandhinagar, Gujarat, 382428, India}
\author{Surabhi Jaiswal} 
\email{szj0071@auburn.edu}
\affiliation{Physics department, Leach science center, Auburn University, Auburn, AL, 36849, USA}
\affiliation{Institute for Plasma Research, HBNI, Gandhinagar, Gujarat, 382428, India}
\author {Manish K Shukla}
\email{shuklamanish786@gmail.com}
\affiliation{Department of Physics and Astrophysics, University of Delhi, Delhi - 110007, India}
\author {Ammar Hakim}
\email{ahakim@pppl.gov}
\affiliation{Princeton Plasma Physics Laboratory, Princeton, New Jersey, 08540, USA}
\author{Edward Thomas} 
\email{etjr@auburn.edu}
\affiliation{Physics department, Leach science center, Auburn University, Auburn, AL, 36849, USA}

\begin{abstract}
A new method called `Configurational Temperature' is introduced in the context of dusty plasma, where the temperature of the dust particles, submerged in the plasma, can be measured directly from the positional information of the individual dust particles and the interaction potential between the dust grains. This method does not require the velocity information of individual particles which is a key parameter to measure the dust temperature in the conventional method. %We do not need the information of velocity of individual dust particles to measure the temperature in this technique.
The technique is initially tested using two dimensional OpenMP parallel Molecular Dynamics and Monte-Carlo simulation and then compared with the temperature evaluating from the experimental data. The experiments have been carried out in Dusty plasma experimental (DPEx) device where a two dimensional stationary plasma crystal of melamine formaldehyde particles is formed in the cathode sheath of a DC glow discharge argon plasma. The dust kinetic temperature is calculated using standard PIV technique at different pressures. %The results are comparable when the dust cloud arranged into crystalline state.
The simulation results matches well with the experimental data at relatively higher pressures where the dust particles arranged into crystalline state or in a strongly coupled fluid state. 
 An extended simulation results for three dimensional case is also presented which can be employed for the temperature measurement of three dimensional dust crystal in laboratory devices.
 %The technique is initially tested using two and three dimensional OpenMP parallel Molecular Dynamics and Monte-Carlo simulation and %finally experimantal data from DPEx device is fed into the new diagnostics and the results are compared with the standard PIV technique.
\end{abstract}

\maketitle

%=========================================================================================

\section{Introduction}
In the presence of micron-sized dust particles within a regular plasma, the electrons get accumulated on the surface of the dust particles due to their high mobility, compared to ions of the background plasma. Once the electrons get adsorbed on the dust particles, the ions flow towards the dust particles. After reaching a steady state of the electron as well as the ion flow the initial excess amount of electrons continues to exist and thus the dust particles get highly negatively charged and the interaction between the dust particles becomes very strong. Not only that, the dust particles due to their low mobility, responds in a very large time scale compared to the electrons as well as the ions of the background plasma. Hence whenever the negatively charged dust particles move, the distribution of the electrons and ions around the dust particles can be considered as a Boltzmannian distribution. \par
In laboratory plasma and other commonly observed plasmas, the interparticle potential between the two dust grains are screened by the background plasma and hence we can think of the interaction potential between the dust particles as a screened Coulomb potential or the Yukawa potential of the form $\phi_d = \frac{Q_d}{4 \pi \epsilon_0 r} e^{-\frac{r}{\lambda_d}}$ where $r$ is the interparticle distance and $\lambda_d$ is the Debye screening length $= \left(\frac{q_i^2 n_i}{\epsilon_0 k_B T_i} + \frac{e^2 n_e}{\epsilon_0 k_B T_e} \right)^{-\frac{1}{2}}$, $Q_d$, $q_i$ and e is the charge of the dust, ion and electron respectively, $T_i$ and $T_e$ is the ion and electron temperature and $n_i$ and $n_e$ is the ion and electron density respectively.\par
Here we have described a method where we utilized the approximate interaction potential between the two dust grains and their positional information to calculate the temperature of dust particles so called \lq\lq configurational temperature". This technique does not rely on the information of velocity of individual dust particles to measure the temperature unlike the other standard methods documented in past \cite{williams:2007, melzer:1996} where the kinetic dust temperature is derived from the root-mean square of the particle velocities v. In the experiments, the dust particles velocity is either calculated by tracking the particles from one frame to the next or by using particle image velocimetry technique which is generally challenging due to the need of very sophisticated hardware arrangement. Therefore configuration temperature can be used as a good alternate to gather the information of dust temperature. 
%In this paper we describe a method to measure the temperature of such dust particles immersed in the plasma only using their positional information and the approximate interaction potential between the two dust grains. In this technique we do not need the information of velocity of individual dust particles to measure the temperature unlike the other standard methods available today.
%=========================================================================================
\section{Configurational Temperature}
The concept of configurational temperature has been widely used by fluid community. Hans Henrik Rugh introduced the idea for the first time in his pioneering paper \cite{rugh:1997}. Later Butler {\it et al} \cite{butler:1998} implemented the algorithm for Monte-Carlo simulation. After that, this diagnostics has been widely used and found to hold good in several critical tests \cite{branka:2010, travis:2008, delhommelle:2002, chialvo:2001, delhommelle:2001}. Of late, the technique is applied to construct thermostats in Molecular Dynamics simulations also \cite{braga:2005, patra:2015, samoletov:2010, gupta:2016}. Very recently, using the same concept, configurational entropy has also been calculated \cite{preisler:2016}. This technique has been found to work well for Monte-Carlo simulation for complex plasmas \cite{rupak:2016} as well. Of late, this technique has been implemented in dusty plasma systems by Himpel {\it et al} \cite{himpel:2019} and compared with experimental data.\\

Below we give a brief introduction of `Configurational Temperature' following the seminal paper by H H Rugh \cite{rugh:1997}. Given any pair potential $\phi(r_{ij})$ that depends only on the interparticle distance, the force on a single particle ($i$) due to all other particles ($j$) can be calculated. The expression of force on the $i^{th}$ particle will be given by equation (\ref{force}). From equation (\ref{config_t}) the `Configurational Temperature' of the system ($T_{conf}$) can be determined in a statistical sense.
\begin{eqnarray}
&& \label{force} F_i = -\nabla_i \sum \limits_{j \neq i}^N \phi(r_{ij})\\
&& \label{config_t} K_B T_{conf} = \frac{\langle \sum \limits_{i=1}^N F_i^2 \rangle}{\langle -\sum \limits _{i=1}^N \nabla_i F_i \rangle}
\end{eqnarray}
where, the angular brackets represent the space time average. \\

Here in this paper, we assume the interparticle interaction between the two dust grains to be Yukawa \cite{konopka:2000}, having the form $\phi(r_{ij}) = \frac{1}{r_{ij}} e^{-\kappa r_{ij}}$ and from experimental parameters determine the value of $\kappa$. Thus in this technique we do not need any information about the velocity of the dust particles to determine the temperature as is done in other standard techniques. We first check the implementation of such diagnostics in our well-benchmarked two-dimensional OpenMP parallel Molecular Dynamics code where we perform extensive runs with both periodic as well as reflecting boundary conditions. The numerical results are described in details in the following section (Sec.~\ref{md}). Further we test the implementation in our two-dimensional OpenMP parallel Monte-Carlo code where, we give a thermal quench to the dusty plasma system and follow the temperature profile of the dusts from the configurational data of the code (Sec.~\ref{mc}). 
 Next, we implemented our \lq Configurational Temperature' diagnostics on the data obtained from dusty plasma experimental (DPEx) device \cite{jaiswal:2015a} and compared the results with the temperature calculated from velocity distribution of the dust particle analyzed using Particle-Image- velocimetry (PIV) technique as described in Sec.~\ref{exp_sec}. Finally we extend our numerical search to three dimensions and carry out numerical simulation using three dimensional OpenMP parallel Molecular Dynamics and Monte-Carlo technique to compare the `Configurational Temperature' obtained from both the simulations independently (Sec.~\ref{3D}). A brief concluding remark is made in Sec.~\ref{conclusion}.

%=========================================================================================

\section{Benchmarking with Molecular Dynamics\label{md}}

The `Configurational Temperature' is obtained from the position snapshots obtained from equilibrium  Molecular Dynamics (MD) simulation and is compared with kinetic temperature obtained directly from MD simulation. The interparticle interaction is taken to be screened Coulomb (or Yukawa) i.e.,
$$\phi(r)= \frac{Q}{4\pi \epsilon_0} \frac{\exp(- r/\lambda_D)}{r},$$
where $r$ is interparticle distance between two particles, $\lambda_D$ is Debye screening length and $Q$ is charge of the particle.\\

We have used normalized units for MD simulations. The length is normalized by mean interpaticle distance $a$, which, in 2D, is related to areal number density by relation $\pi a^2 n_d=1$. Time is normalized by $\Omega^{-1}_{pd}$ where $\Omega^2_{pd} = (Q^2/4\epsilon_0 m \bar n)$ and $\bar n$ is normalized number density in 2D. All energy measurements are carried out in units of $Q^2/4\pi\epsilon_0 a$. Therefore, in normalized units, the equation of motion of $i^{th}$ particle is written as 
\begin{equation}
 \frac{d^2 \vec r_{i}}{dt^2}=\vec f_{i}= \sum_{j,j\neq i}^{N} \left(1+ \kappa \; r_{ij} \right) \frac{e^{-\kappa r_{ij}}}{r_{ij}^3} \vec r_{ij}
 \label{MD_eq_1},
\end{equation}
where $\kappa~(=a/\lambda_D)$ is screening parameter. Eq. (\ref{MD_eq_1}) is integrated using Leapfrog algorithm. For a fixed time step $dt$, the position and velocities at a time step $(n+1)$ are given by
\begin{eqnarray*}
&& r_{n+1}= r_n+ v_n dt+\frac{1}{2} f_n dt^2\\
&& v_{n+1}= v_n+ \frac {1}{2} f_n dt+\frac{1}{2} f_{n+1} dt.
\end{eqnarray*}
Temperature ($k_BT$) is measured in units of $Q^2/4\pi\epsilon_0 a$, and since each translational degree of freedom contributes to $k_BT/2$ to the kinetic energy, the temperature of our 2D system is  
\begin{equation}
 T = \frac{1}{2N}\sum_{i=1}^{N} v_i^2.
\end{equation}
The thermodynamic state of Yukawa system can  be  described completely either by taking number density and temperature as free parameters \cite{manish:2017} or  by using  dimensionless parameters, (1) $\kappa$, defined earlier and (2)  $\Gamma$ $(= Q^2/4\pi\epsilon_0 k_BT)$ as free parameters\cite{hamaguchi:1997}. We have chosen $(\Gamma, \kappa)$ to characterize our system. The parameter $\kappa$ controls the interaction potential for example,  $\kappa \rightarrow \infty$ refers to an ideal gas while $\kappa \rightarrow 0$ refers to a Coulomb gas. The strong coupling parameter $\Gamma^{*}= \Gamma \exp(-\kappa)$ which is the ratio of the mean inter-particle potential energy to the mean kinetic energy, is used as a measure of coupling strength in dusty plasmas. For $\Gamma^* \ll 1$, Yukawa system behaves like an ideal gas, $\Gamma^* \sim 1$ corresponds to an interacting fluid whereas $\Gamma^* \gg 1$  refers to a condensed solid state.  In our chosen normalization, the coupling parameter $\Gamma$ becomes the inverse of normalized dust temperature  i.e. $\Gamma= 1/T$.\\

To obtain a desired temperature in our MD simulation, we have used Berendsen thermostat. The system is kept in contact with the bath for first $10^5$ time steps and then isolated for next $10^5$ steps. To check the versatility of `Configurational Temperature', we have taken data with (i) varying number of particles from $N=500$ to $N=5000$ while keeping $\Gamma$ and $\kappa$ constant, (ii) with   varying $\Gamma$ ranging from $10$ to $500$ (where $\Gamma_{crit} = 170$ being the critical point at $\kappa = 1$) with $N$ and $\kappa$ fixed, and, (iii) with varying $\kappa$ ranging from $0.5$ to $4.0$ with fixed $N$ and $\Gamma$. For each of the three cases we have also taken data in periodic boundary condition as well as in perfectly reflecting  boundary conditions. 

%======================================================================================================================
 
\begin{figure}[h!]
\includegraphics[scale=0.65]{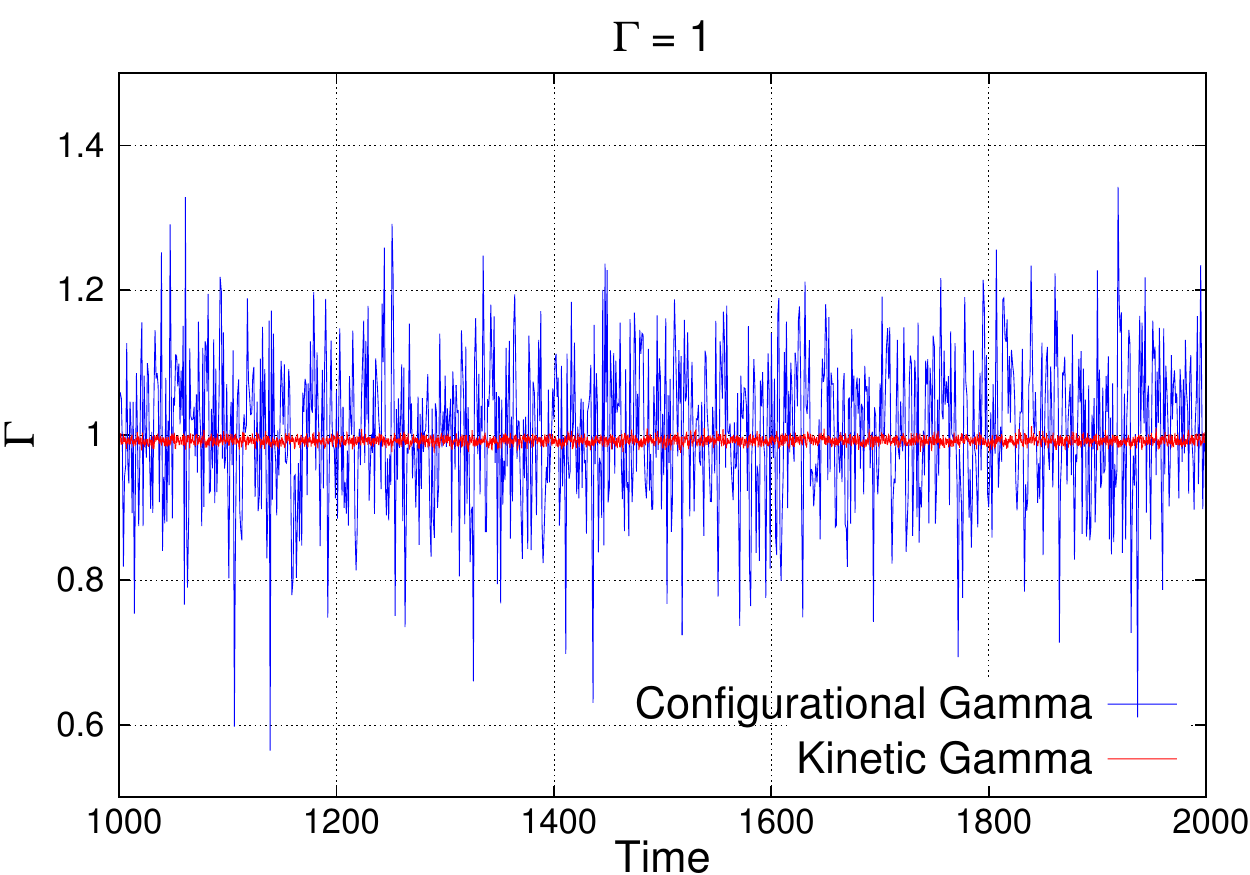}
\caption{Comparison between Configurational $\Gamma$ and Kinetic $\Gamma$ at the micro-canonical run at $N = 5000$, $\kappa = 1$ and $\Gamma = 1$ with periodic boundary condition.}
\label{Gamma_1}
\end{figure}

\begin{figure}[h!]
\includegraphics[scale=0.65]{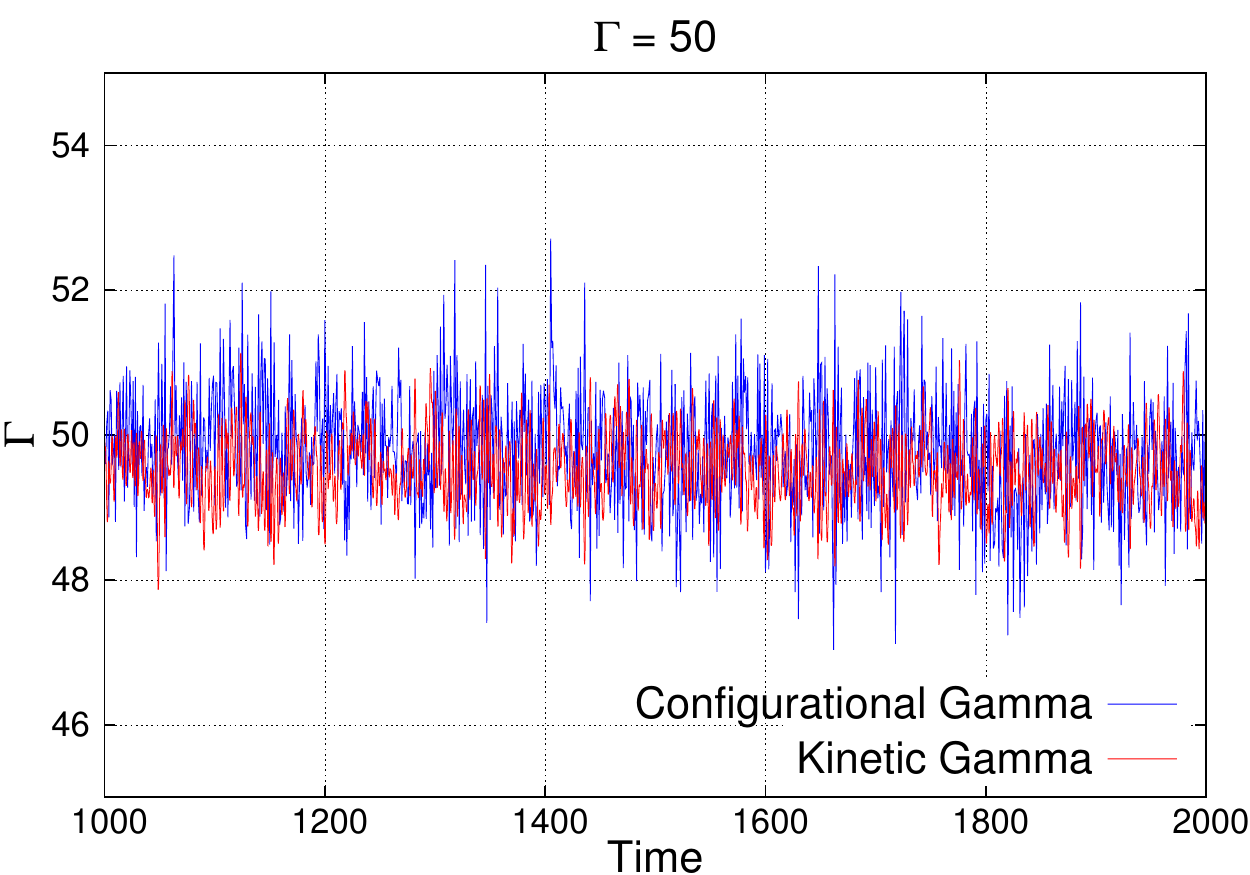}
\caption{Comparison between Configurational $\Gamma$ and Kinetic $\Gamma$ at the micro-canonical run at $N = 5000$, $\kappa = 1$ and $\Gamma = 50$ with periodic boundary condition.}
\label{Gamma_50}
\end{figure}

\begin{figure}[h!]
\includegraphics[scale=0.65]{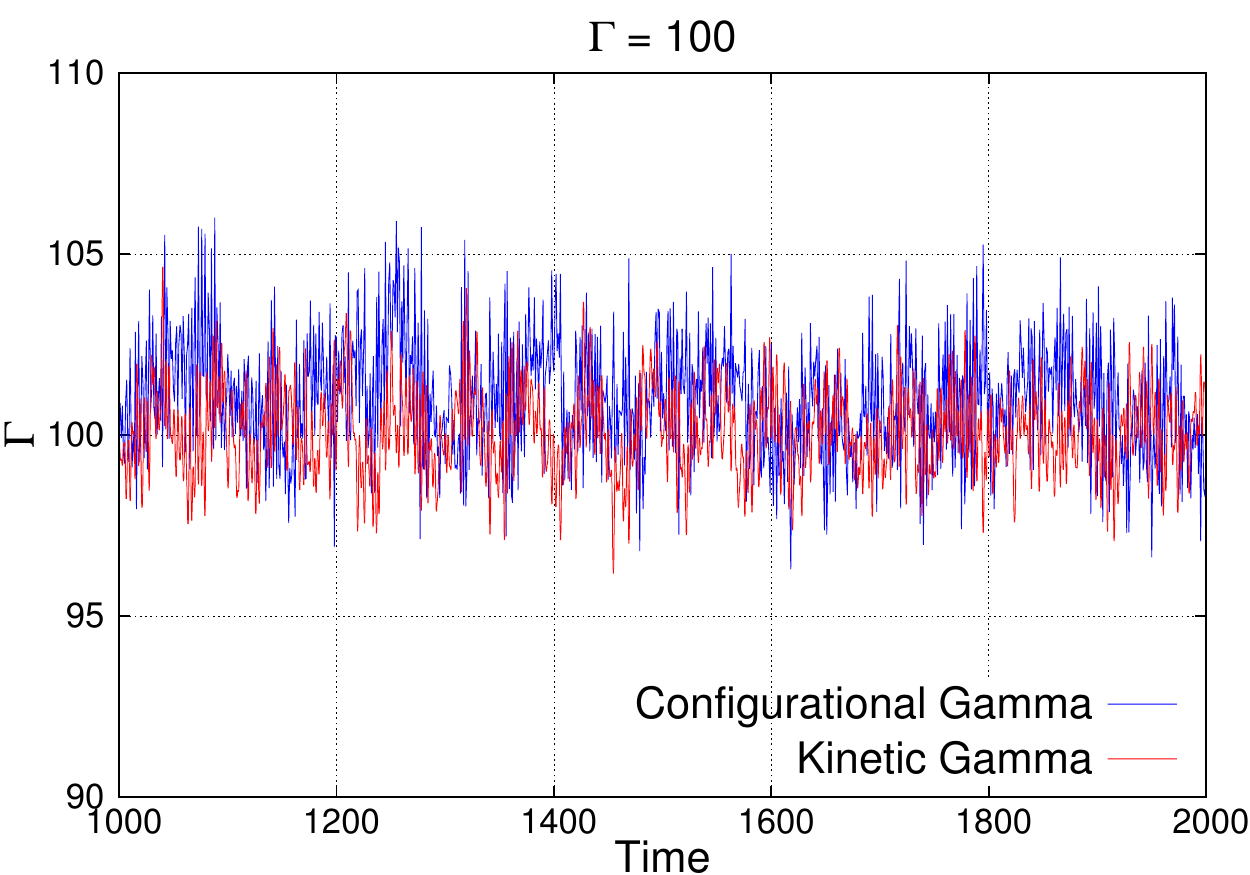}
\caption{Comparison between Configurational $\Gamma$ and Kinetic $\Gamma$ at the micro-canonical run at $N = 5000$, $\kappa = 1$ and $\Gamma = 100$ with periodic boundary condition.}
\label{Gamma_100}
\end{figure}

\begin{figure}[h!]
\includegraphics[scale=0.65]{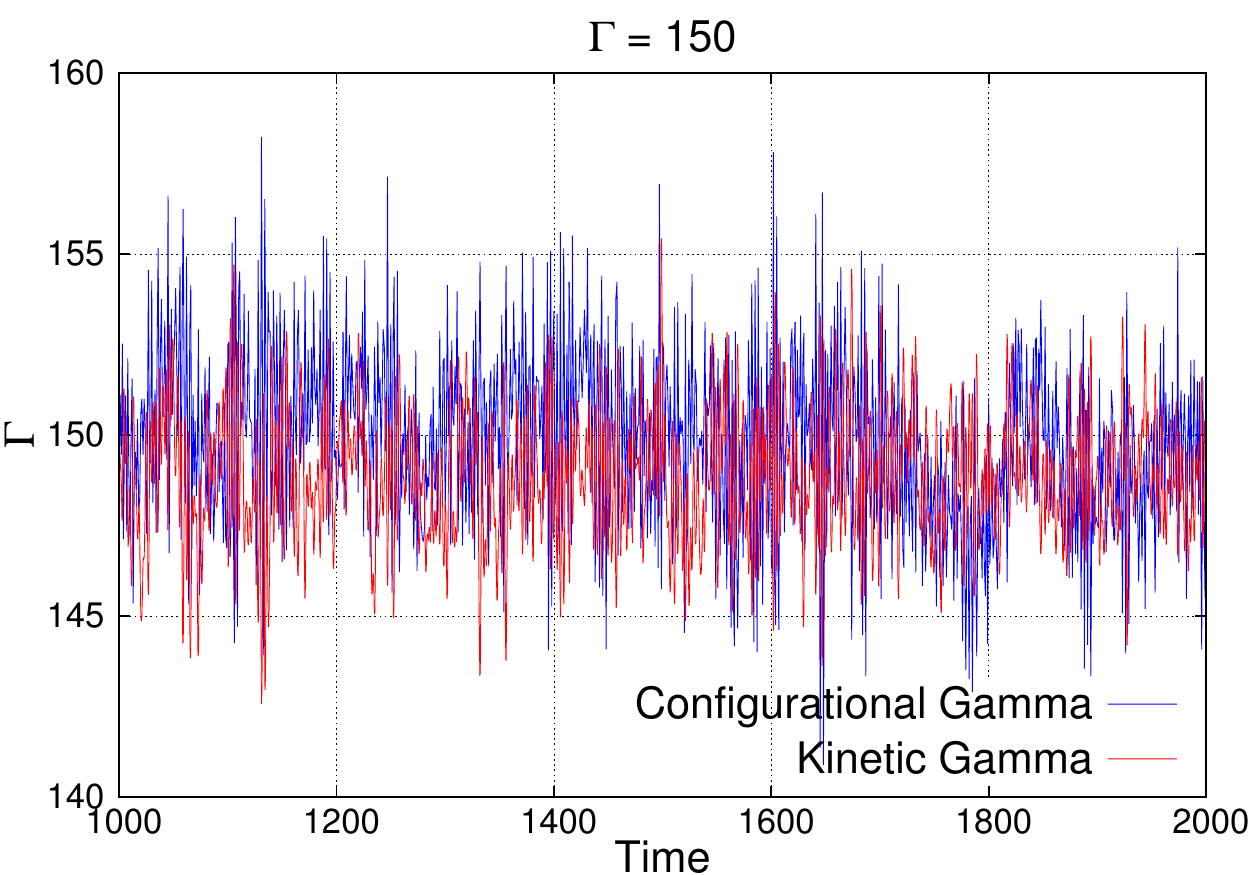}
\caption{Comparison between Configurational $\Gamma$ and Kinetic $\Gamma$ at the micro-canonical run at $N = 5000$, $\kappa = 1$ and $\Gamma = 150$ with periodic boundary condition.}
\label{Gamma_150}
\end{figure}

\begin{figure}[h!]
\includegraphics[scale=0.65]{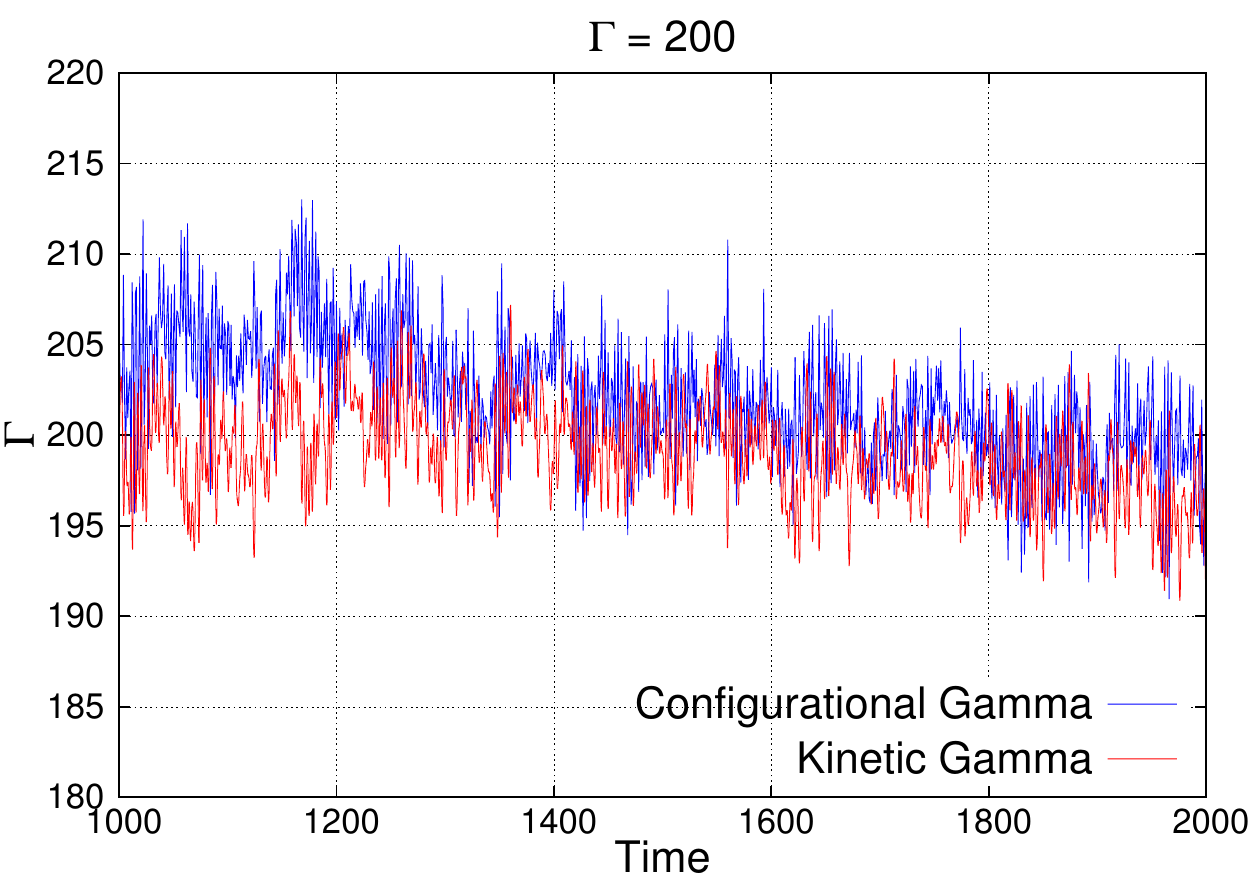}
\caption{Comparison between Configurational $\Gamma$ and Kinetic $\Gamma$ at the micro-canonical run at $N = 5000$, $\kappa = 1$ and $\Gamma = 200$ with periodic boundary condition.}
\label{Gamma_200}
\end{figure}

\begin{figure}[h!]
\includegraphics[scale=0.65]{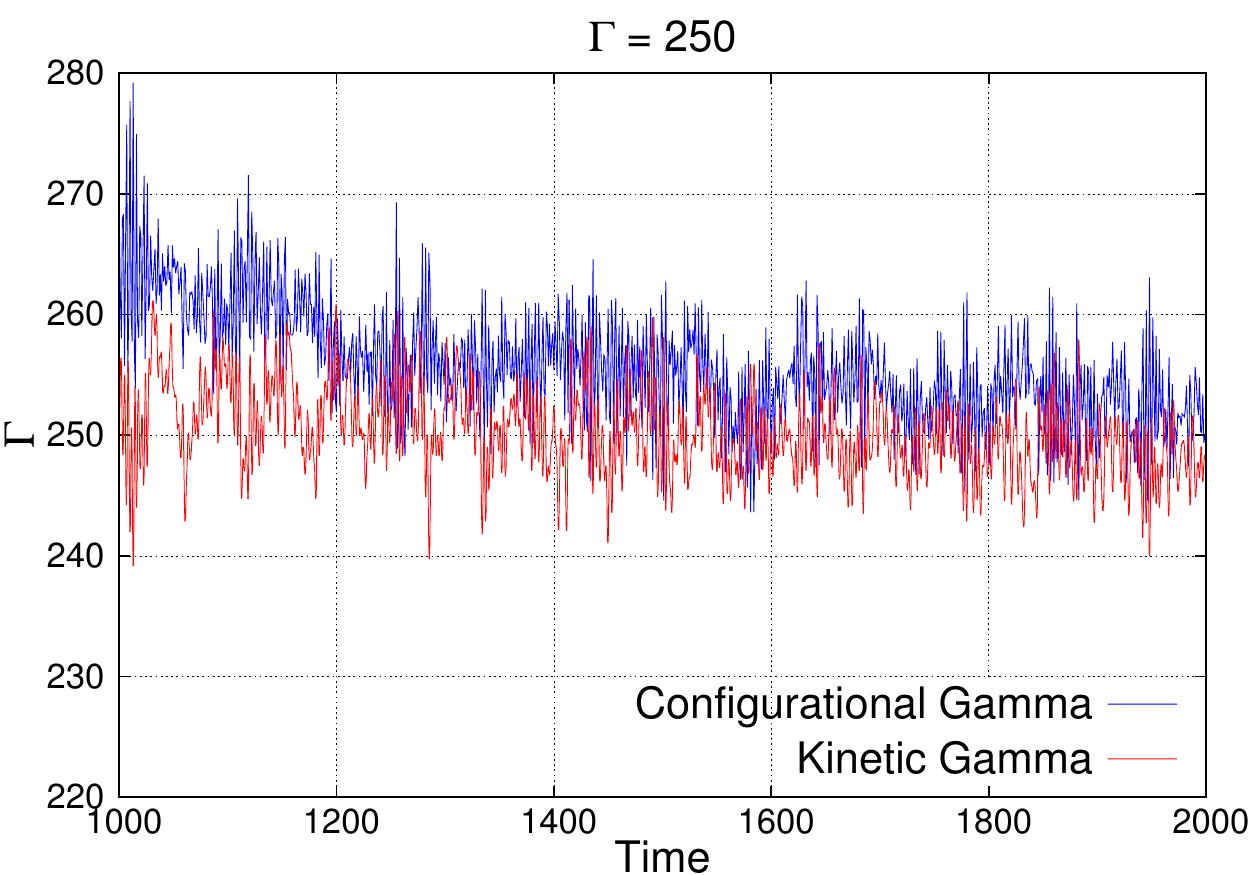}
\caption{Comparison between Configurational $\Gamma$ and Kinetic $\Gamma$ at the micro-canonical run at $N = 5000$, $\kappa = 1$ and $\Gamma = 250$ with periodic boundary condition.}
\label{Gamma_250}
\end{figure}

%==============================================

\begin{figure}[h!]
\includegraphics[scale=0.65]{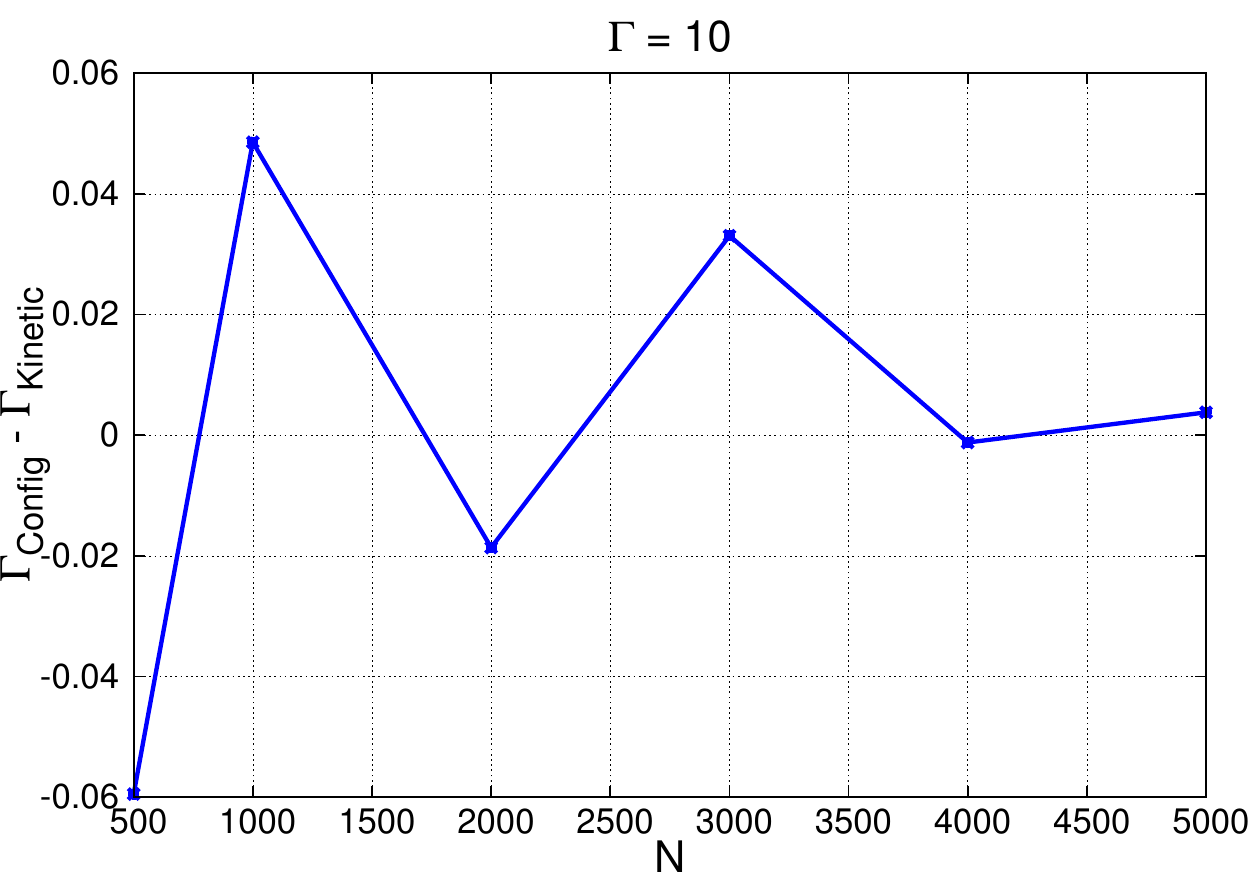}
\caption{Relative error between Configurational $\Gamma$ and Kinetic $\Gamma$ at $\Gamma = 10$ and $\kappa = 1$ with periodic boundary condition. The $\Gamma$ evaluated from both techniques appear to be convergent as $N$ increases.}
\label{N_Comp_periodic}
\end{figure}

\begin{figure}[h!]
\includegraphics[scale=0.65]{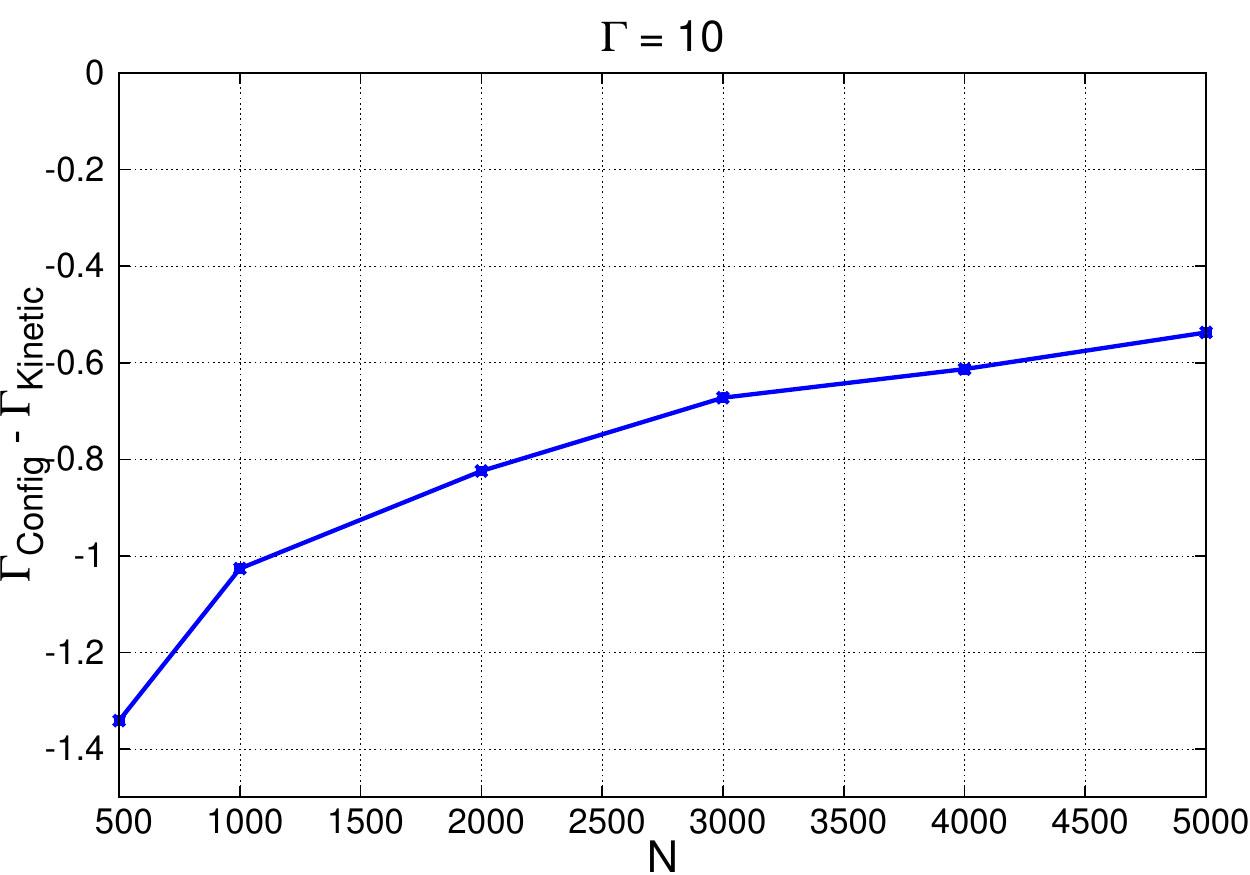}
\caption{Relative error between Configurational $\Gamma$ and Kinetic $\Gamma$ at $\Gamma = 10$ and $\kappa = 1$ with reflecting boundary condition. The $\Gamma$ evaluated from both techniques appear to be convergent as $N$ increases.}
\label{N_Comp_reflecting}
\end{figure}

%==============================================

\begin{figure}[h!]
\includegraphics[scale=0.65]{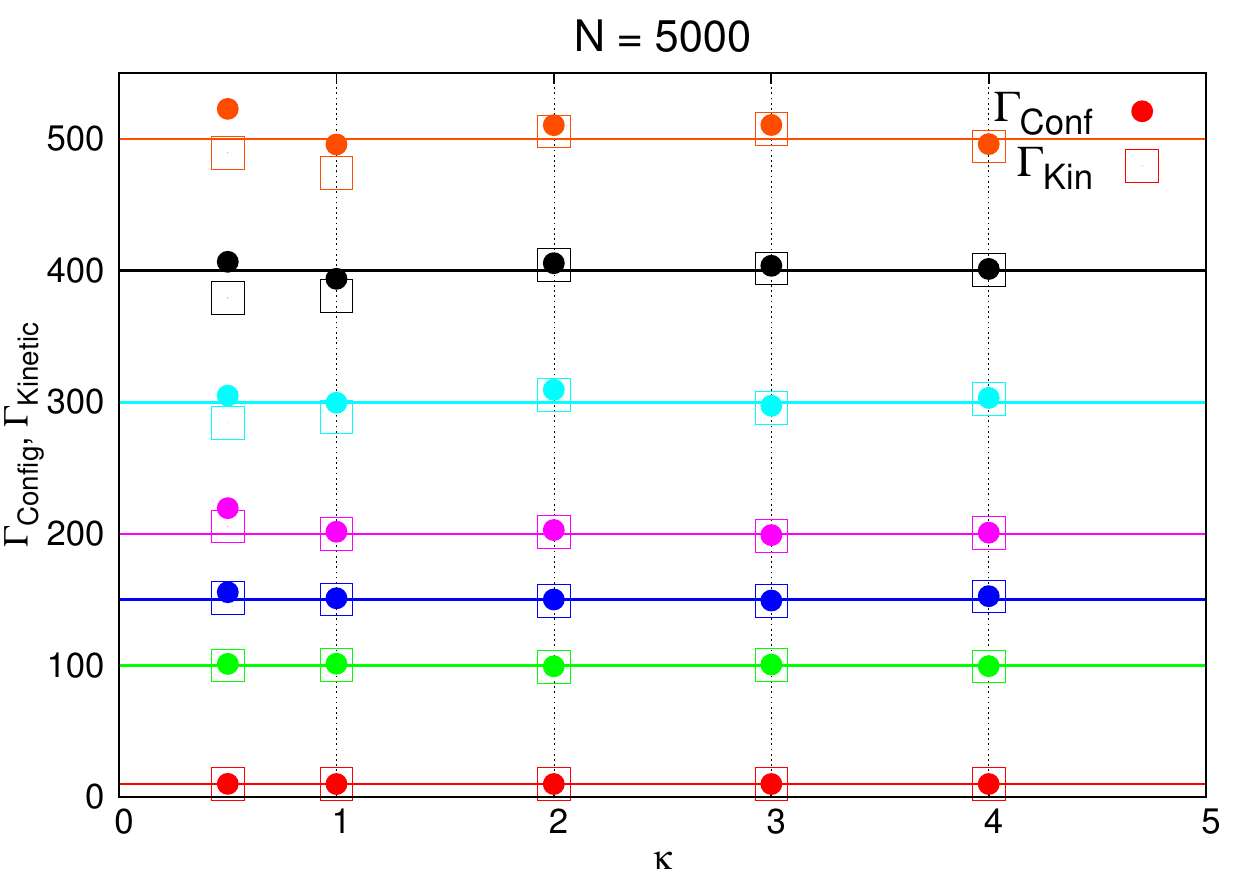}
\caption{Configurational $\Gamma$ and Kinetic $\Gamma$ with $\Gamma$ ranging from $0$ to $500$ and $\kappa$ varying between $0.5$ to $4$ with $N = 5000$ and reflecting boundary condition. The $\Gamma$ evaluated from both techniques appear to be convergent for all values.}
\label{kappa_vary_periodic}
\end{figure}

\begin{figure}[h!]
\includegraphics[scale=0.65]{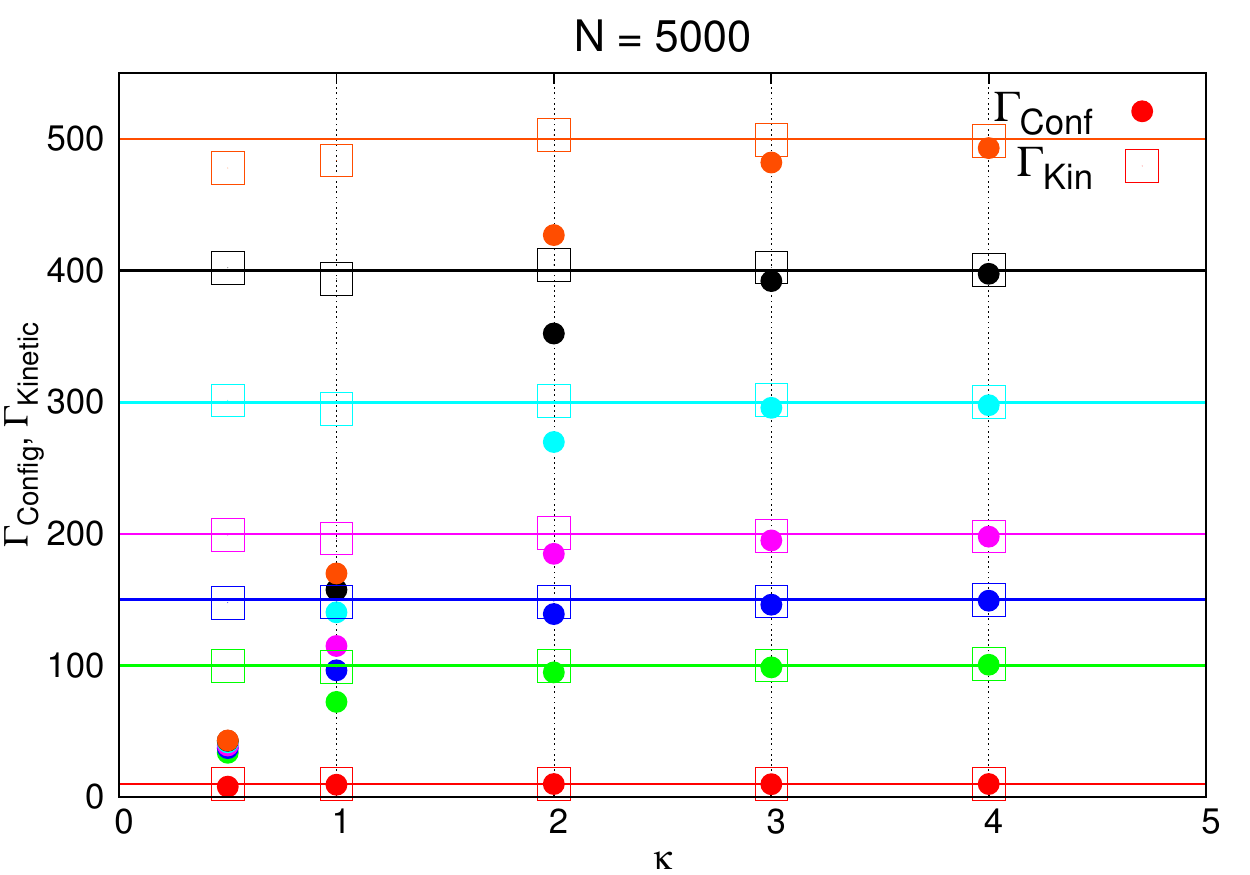}
\caption{Configurational $\Gamma$ and Kinetic $\Gamma$ with $\Gamma$ ranging from $0$ to $500$ and $\kappa$ varying between $0.5$ to $4$ with $N = 5000$ and reflecting boundary condition. The $\Gamma$ evaluated from both techniques appear to be convergent as $\kappa$ increases.}
\label{kappa_vary_reflecting}
\end{figure}

%=========================================================================================

\section{Benchmarking with Monte-Carlo Simulation\label{mc}}

%==============================================

\subsection{Monte-Carlo Algorithm:}

The expectation value of a statistical quantity $Q(\bar{x})$ is given by,
\begin{eqnarray*}
\textless Q(\bar{x}) \textgreater = \frac{\int d\bar{x} Q(\bar{x})e^{-\beta H(\bar{x})}}{\int d\bar{x}e^{-\beta H(\bar{x})}}
\end{eqnarray*}
In the discretised form it can be written as,
\begin{eqnarray*}
\textless Q(\bar{x}) \textgreater = \frac{\sum \limits_{i=1}^M Q(\bar{x}_i)e^{- \beta H(\bar{x}_i)}}{\sum \limits_{i=1}^M e^{- \beta H(\bar{x}_i)}}
\end{eqnarray*}
where $\bar{x}$ ranges over the full phase space. So for $N (\rightarrow \infty)$  particles the phase space dimension will be $4N$ which might be computationally very costly. So we generate a function $p(\bar{x_i})$ such that we mostly concentrate on that part of the phase space from where the contribution to the integral will be larger. This technique is called the Importance sampling method. 
\begin{eqnarray*}
\textless Q(\bar{x}) \textgreater = \frac{\sum \limits_{i=1}^M \frac{Q(\bar{x}_i)e^{- \beta H(\bar{x}_i)}}{p(x_i)}}{\sum \limits_{i=1}^M \frac{e^{- \beta H(\bar{x}_i)}}{p(x_i)}}
\end{eqnarray*}
We choose $P(\bar{x}_i) = e^{- \beta H(\bar{x}_i)}$. The equation reduces to the form,
\begin{eqnarray*}
\textless Q(\bar{x}^*) \textgreater = \frac{\sum \limits_{i=1}^M Q(\bar{x}_i^*)}{M}
\end{eqnarray*} 
where $\bar{x}_i^*$ are the specially chosen points of the phase space that contributes largely to the integral. It is expected that $\textless Q(\bar{x}^*) \textgreater \rightarrow \textless Q(\bar{x}) \textgreater$ for large Monte-Carlo Steps. The concept of the ``specially chosen points" is realised through the Metropolis Algorithm\cite{metropolis:1953}. Instead of choosing successive states $x_i$, we construct a Markov process where each state $x_{i+1}$, is generated from its previous state $x_i$ with a suitable transition probability $W(x_i \rightarrow x_{i+1})$. Thus the problem boils down to generate the transition probability. The technique is described below.\\
The Principle of Detailed Balance is written as:
\begin{eqnarray*}
&& \frac{\partial P(x_i)(t)}{\partial t} = - [ P(x_i)(t) W(x_i \rightarrow x_{i+1}) \\
&& ~~~~~~~~~~~~~~~~~ - P(x_{i+1})(t) W(x_{i+1} \rightarrow x_i)]
\end{eqnarray*}
Under Equilibrium condition the equation looks as;
\begin{eqnarray*}
&& P_{eq}(x_i) W(x_i \rightarrow x_{i+1}) = P_{eq}(x_{i+1}) W(x_{i+1} \rightarrow x_i)\\
\Rightarrow && \frac{W(x_i \rightarrow x_{i+1})}{W(x_{i+1} \rightarrow x_i)} = \frac{P_{eq}(x_{i+1})}{P_{eq}(x_i)}\\
\Rightarrow && \frac{W(x_i \rightarrow x_{i+1})}{W(x_{i+1} \rightarrow x_i)} = e^{- \beta \delta H}
\end{eqnarray*}
where $\delta H = H(x_{i+1}) - H(x_i)$.\\
In order to remove the arbitrariness in the choice of $W$, we choose 
\begin{eqnarray*}
W(x_i \rightarrow x_{i+1}) &=& e^{-\beta \delta H}, ~ {\text {if}} ~~ \delta H \textgreater 0\\
&=& 1, ~~~~~~~ {\text {otherwise}}
\end{eqnarray*}
Thus for a statistical system the energetically favoured states are immidiately accepted. But the energetically unfavoured states also do have a probability to get accepted.

%==============================================

\subsection{Simulation Technique:}

We apply the Monte-Carlo algorithm to simulate a classical strongly interacting multiparticle canonical ensemble to study the different phases of matter. The system we have considered has been described above. We consider  Yukawa interaction potential. We start with a random initial configuration and the total energy of the system is calculated. For  every $i^{th}$ particle we calculate the potential energy as 
\begin{equation}
\phi_i = \sum\limits_{j=1}^N \frac{1}{r_{ij}} e^ {-\kappa r_{ij}}
\end{equation} 
with $i \neq j$. Thus we make a sum of the total energy of all the particles and take a mean to calculate the mean energy per particle.\\

Then we randomly move a single particle (say the $i^{th}$ particle) within a predefined maximum amplitude and calculate the total energy of the system again. Now take the energy difference of the mean energy per particle and calculate the transition probability as 
\begin{equation}
e^{- (W_{moved} - W)/T}
\end{equation} 
where $W$ is the total energy of the system. Then we generate a random number between $0$ and $1$ (using the standard fortran random number generator) and compare the magnitude of the two. If the transition probability is higher than the random number we can say that the event of the movement of the $i^{th}$ particle with that amount is more probable than a random event. Hence we accept the move. If the transition probability is lesser than the random number then we reject that move and the particle is then given back its old position. We keep our acceptance ratio in between $0.3$ to $0.5$ to make the importance sampling efficient.\\

Now in order to reduce the computational cost, before the movement we store the energy of every particle in an array and create another array to store the energy for every particle after the move and take the difference of the arrays only during the computation of the transition probability because for the random movement of the $i^{th}$ particle from its initial position the interaction energy changes for the  $i^{th}$ particle only and there will be no change into the interaction energy of any other particle. Hence we dont need to calculate the average interaction energy of all the particles every time to calculate the transition probability for each particle. For example, for the movement of the $i^{th}$ particle, instead of calculating in the previous manner we calculate the transition probability as 
\begin{equation}
e^{- (W_{moved}(i) - W(i))/T}
\end{equation} 
where $W_{i}$ is the energy of the $i^{th}$ particle only. This reduced the computational cost by a great factor.\\

After this we move to another particle and check the same thing with it. Thus we touch all the particles and this is called one Monte-Carlo Step. Eventually it is one of the snapshots of the ensemble or it can also be identified with one of the microstates of the system. After every Monte-Carlo step we calcutate the total energy as well as the positions of all the particles of the new configuration of the system. We also tag all the particles since it helps to calculate the diffusion and other properties of the system though it computationally expensive.\\

Now to generate the other microstates by creating several MC steps. We just continue the displacements and for large number of steps which eventually shows a constant behavior in energy for all the microstates (though there are some initial transient which is due to the mismatch of the position of the particles with the initial temperature).\\

%==============================================

\subsection{Parameter Details:}
We have developed an Open-MP parallel two dimensional multiparticle Monte-Carlo code using the above-mentioned algorithm and benchmark it with the existing molecular dynamic code as well as MPMD-2D code\cite{gupta:2018, charan:2018}. We reproduce the two dimensional pair-correlation function of dusty plasma both in liquid and solid state as previously obtained from both the codes.\\

We run our code with $N = 1225$, $\bar{n} = \frac{1}{\pi}$ and a cut-off radius $R_{cut} = 10 a_0$, where $a_0$ is the Weigner-Seitz cell radius. We choose the average random movement of any particle ($\delta$) in such a way so that the acceptance ratio is guaranteed to remain below $0.5$ and above $0.1$.\\

\begin{figure}[h!]
\includegraphics[scale=0.65]{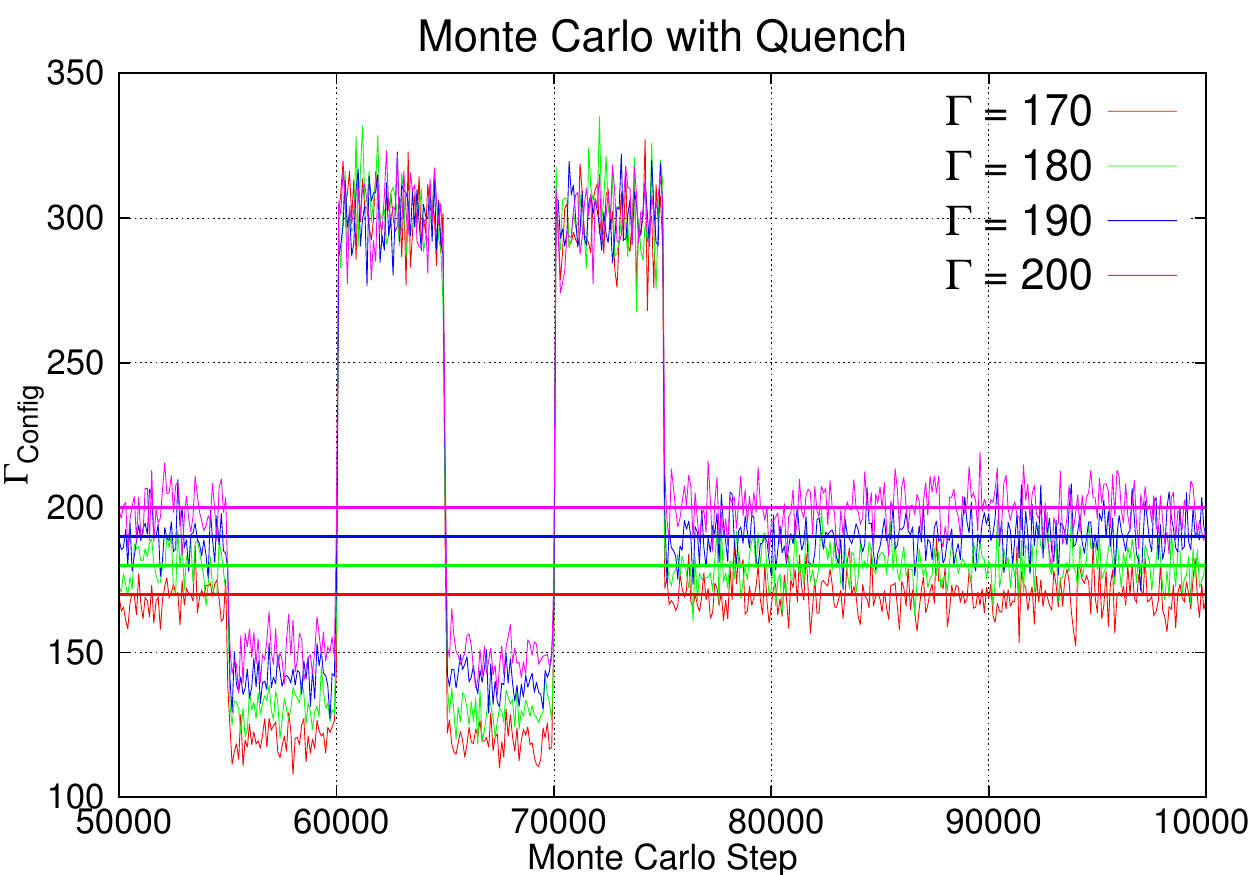}
\caption{Monte-Carlo run with quench with $\Gamma = 170$(red), $180$(green), $190$(blue) and $200$(magenta).}
\label{MC}
\end{figure}

Next, we start with four different $\Gamma$ values (Fig. \ref{MC}) viz. $\Gamma = 170$(red), $180$(green), $190$(blue) and $200$(magenta). We allow the system to evolve upto $50000$ Monte-Carlo step ($MCS$) and get thermalised. At $MCS = 55000$ we suddenly reduce the value of $\Gamma$ by $50$ from its initial value and allow the system to relax for next $5000$ $MCS$. Then at $MCS = 60000$ we give a deep quench to the system and set $\Gamma = 300$. After a relaxation upto $MCS = 65000$ at $\Gamma = 300$, we again set the $\Gamma$ to its previous respective values at $MCS = 55001$. We allow the system to thermalise for $5000$ more $MCS$ and at $MCS = 70000$ we again apply the quench and take all the four cases to $\Gamma = 300$ and allow it to evolve upto $MCS = 75000$. After that, we set back the original $\Gamma$ values to each of them at $MCS = 0$ and evolve further upto $MCS = 100000$.\\

We notice that the `Configurational Temperature' follows the sudden quenches promptly and matches with the externally set temperature quite well. The fluctuations around the expected temperature can be reduced by increasing the number of particles in the code. It is also checked that as we increase the number of particles ($N$), the fluctuation also reduced by $\frac{1}{\sqrt{N}}$.

%=========================================================================================

\section{Benchmarking with Experimental Data \label{exp_sec}} 

\subsection{Description of the experimental device}
\begin{figure}[h!]
\includegraphics[scale=0.25]{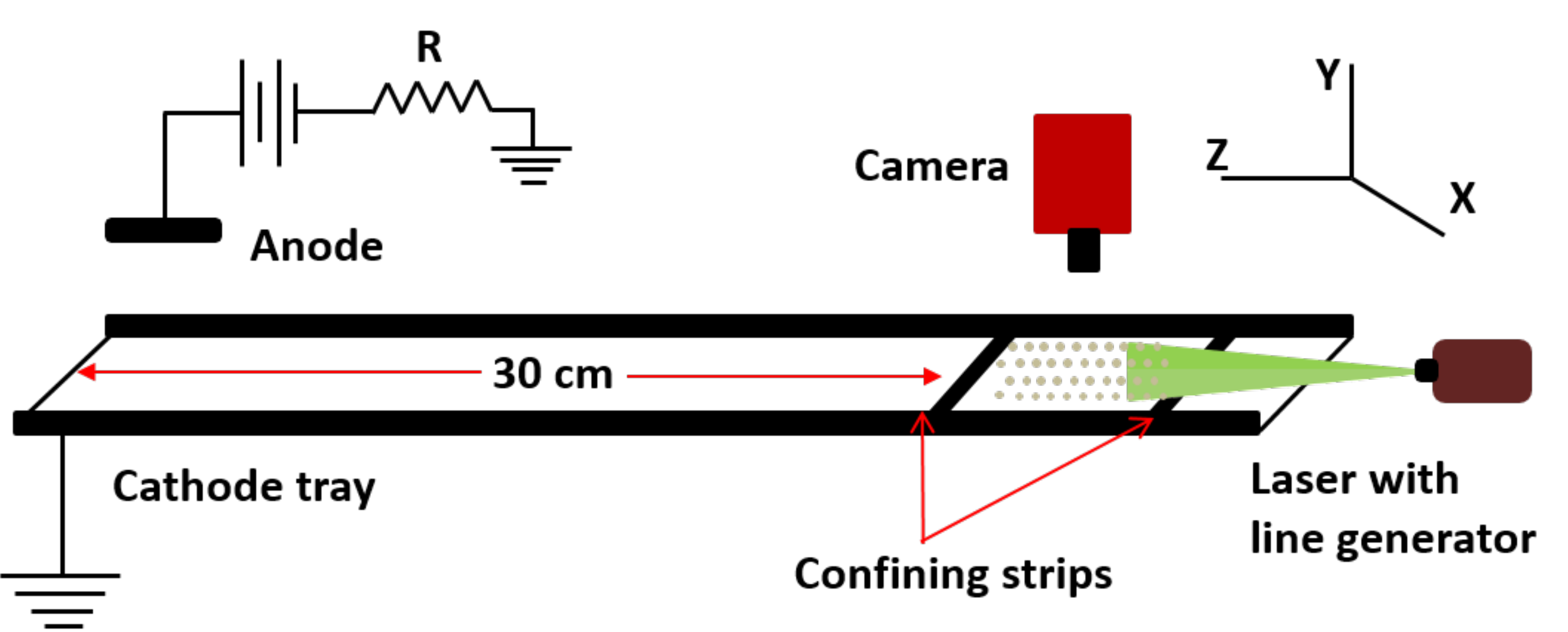}
\caption{experimental arrangement for dusty plasma crystal experiment.}
\label{exp}
\end{figure}

\begin{figure}[h!]
\includegraphics[scale=0.4]{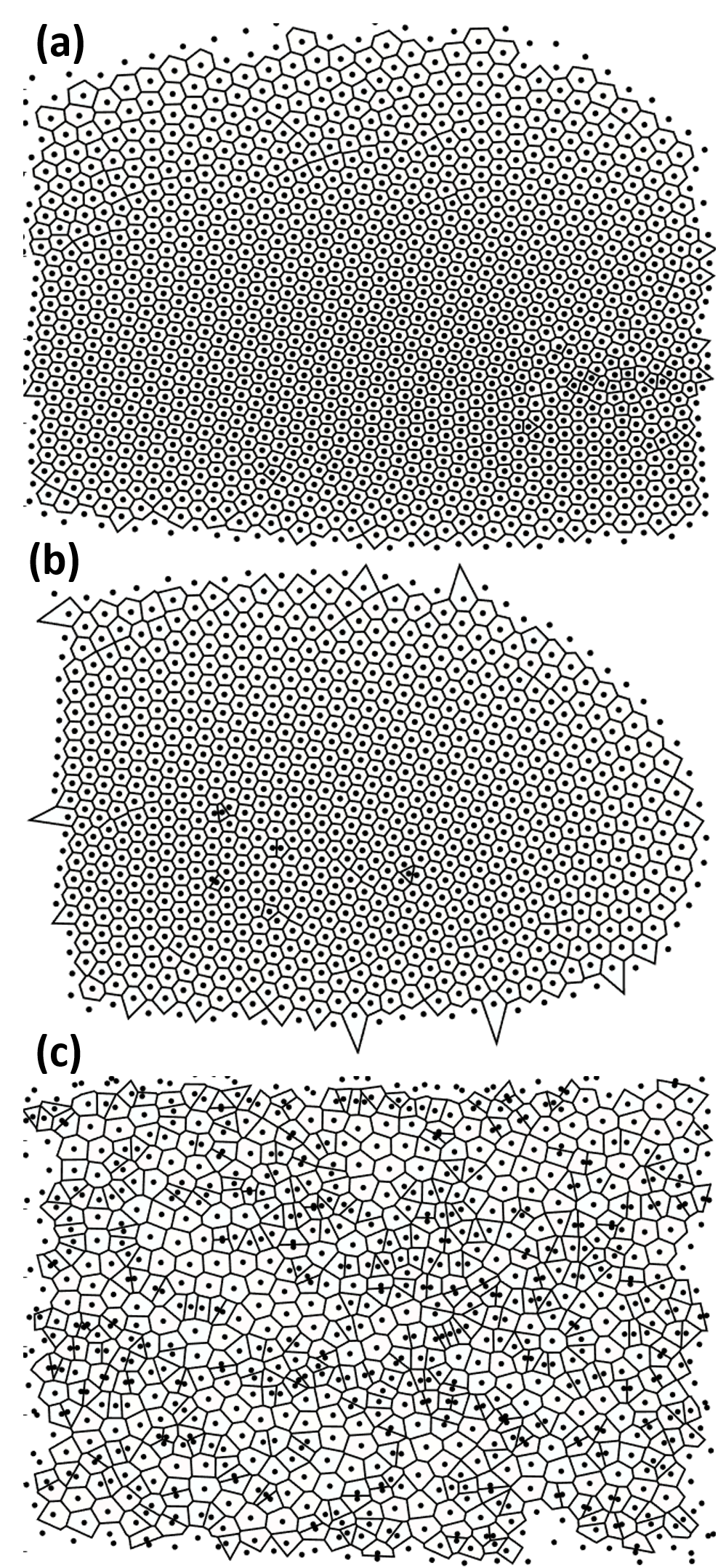}
\caption{Voronoi diagram of the particle cloud for (a) 15 Pa, (b) 13 Pa and (c) 11 Pa.}
\label{voronoi}
\end{figure}

We further implemented our configurational temperature diagnostic on two dimensional finite particle cluster. The measurement has been made in Dusty plasma Experimental (DPEx) device which consists of a {$\Pi-$} shaped pyrex glass tube with a horizontal section of 8 cm inner diameter and 65 cm length. A detailed description of the experimental setup and its operational characteristic is reported elsewhere (see Ref. \cite{jaiswal:2019}). 
A DC glow discharge Argon plasma is strike between a 3 cm disc shaped anode and long grounded cathode tray of 40 cm (see Fig.~\ref{exp}) by applying a discharge voltage in the range $310-320$~Volts. The corresponding discharge current is in the range of $2-5$ mA. Mono-disperse melamine-formaldehyde spheres of diameter $4.38\pm 0.06~\mu$m is used as a dust component. Once introduced into the plasma, these particles become negatively charged by collecting more electrons than ions and trapped in the plasma sheath boundary above the grounded cathode. A couple of stainless steel strip have been used to confine the cloud in the radial and axial direction. The particle cloud is illuminated by a horizontally expanded thin sheet of green laser light (532 nm, 100 mW) which is sufficiently constricted vertically to study an individual
layer of the dust cloud. The Mie-scattered light from the dust particles is captured by a CCD camera (shown in Fig~\ref{exp}) at 25 fps with a resolution of 9 $\mu$m/pixel and the images are stored into a high - speed computer. The particles are arranged into a hexagonal structure at a pressure of 14 Pa. The dust temperature generally depends on discharge parameter. Therefore we vary the background neutral pressure from 15-11 Pa to be able to investigate particle cluster at different temperature. The voronoi diagram of the particle cloud at different pressure is shown in Fig.~\ref{voronoi}. It can be seen from the figure that the particles formed a hexagonal structure at 14 Pa. On reducing the pressure the crystal melts and formed a fluid state at 11 Pa.%\par
%The particles are arranged into a hexagonal structure at a pressure of 14 Pa and become liquid state when the pressure is reduced to 11 Pa. 
%The variation of the gas pressure is known to result in a change of the particle temperature and hence allows to investigate clusters with different temperatures \\
\subsection{Evaluating temperature from experimental data}
\begin{figure}[h!]
\includegraphics[scale=0.65]{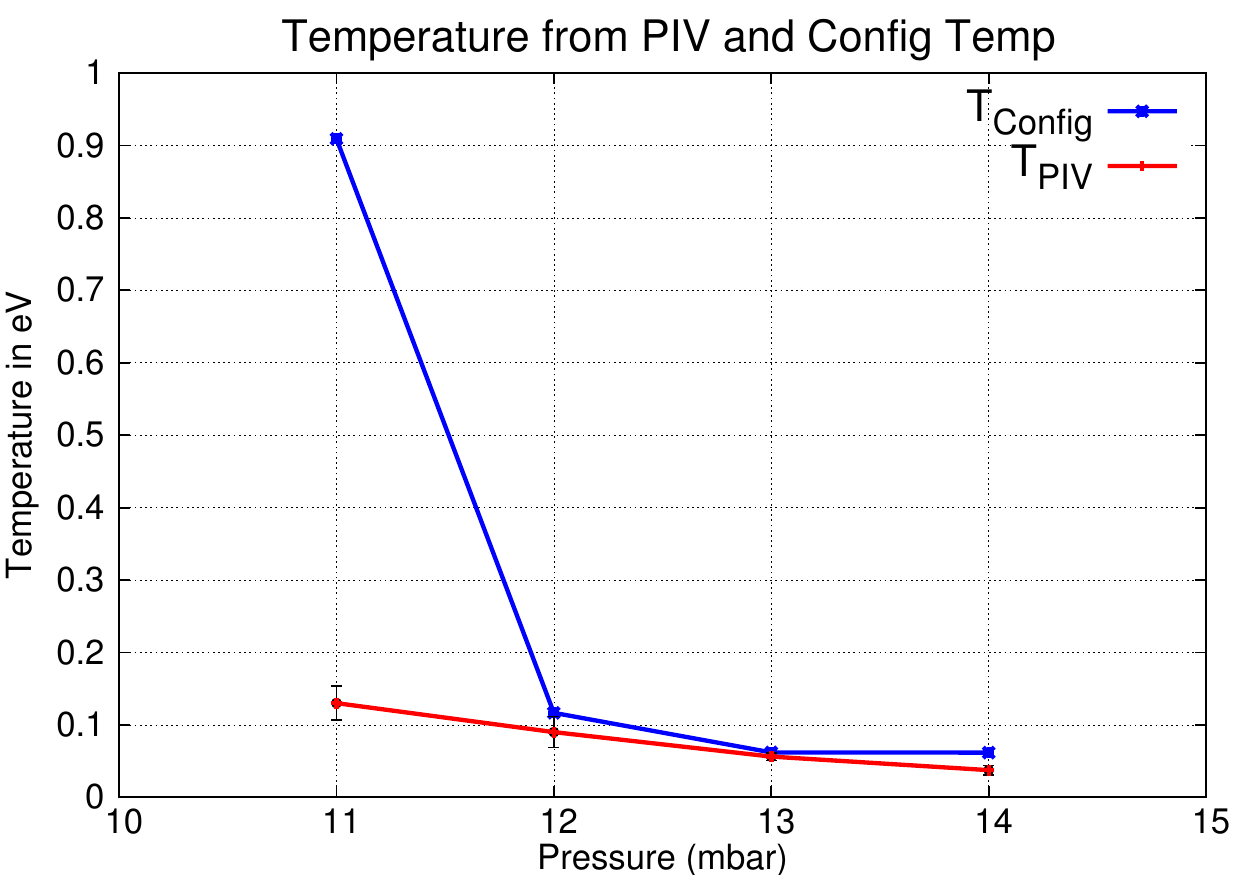}
\caption{Comparison between temperature evaluated from PIV analysis and `Configurational Temperature' for experimental data obtained from DPEx device. At low pressure (fluid state) the `Configurational Temperature' deviates from the temperature evaluated from PIV analysis. However, at high pressure (solid state) the temperature evaluated from both the analysis matches quite well.}
\label{DPEx}
\end{figure}

For the experimental measurement of the kinetic temperature a velocity distribution function in two orthogonal directions has been calculated. For a better  measurement of the distribution function, a video sequence of 100 frames has been chosen where the particle image velocimetry (PIV) analysis package, DAVIS 8 \cite{davis} is used to construct the velocity vector fields. %The distribution function is found to reasonably well fit using a Maxwellian distribution (solid line) on the experimental data points in the figures.
 The kinetic dust temperature is calculated from the width of the measured distribution by using the formula $E =\frac{1}{2}m \langle v^2_{x,y} \rangle = \frac{1}{2}k_B T_{x,y}$. The measured kinetic temperature along with the configuration temperature calculated using our configurational temperature diagnostic is shown in Fig~\ref{DPEx}. The configurational temperature has been derived using 300 particles, for better statistics. It is also worth mentioning that, for the application to real measurement data the charge and screening length is needed. In order to estimate the dust charge, we have followed the procedure outlined in the paper by Khrapak \textit{et al.} \cite{khrapak:2010} that allows estimating the charge based on its dependence on the ratio of the charge density of the particles to that of the ions, defined as Havnes parameter \cite{havnes:1990}  %particle-to-plasma density ratio, defined as Havnes parameter 
\begin{equation}
P = \frac{a T_e}{e^2}\frac{n_p}{n_0}\cong 695 a T_e \frac{n_p}{n_0},    
\end{equation}
where a is the particle radius (in $\mu$m), T$_e$ is the electron temperature (in eV) and n$_p$ and n$_0$ is the particle and plasma densities, respectively. By knowing the information of P, one can estimate the dimensionless charge ($z = e^2 Z_d/4\pi\epsilon_0ak_BT_e$) Khrapak \textit{et al.} \cite{khrapak:2010} and hence the corresponding charge residing on the particle, Q. The determination of the cluster temperature is quite consistent using either the kinetic temperature or the configuration temperature however, the configuration temperature is almost 4 times the kinetic temperature at 11 Pa. This deviation may be arises from screening length or charge variation at lower pressure. Further, the uncertainty in the calculation of the absolute value of particle charges might also have affected the result. A detailed investigation of the configuration temperature at lower pressure may be a worthwhile project to carry out an in-depth exploration of the method. From both the analysis it is clearly seen that at lower pressure the temperature is higher. A rapid fall in the temperature is observed with increasing pressure and above 13 Pa it decays slowly and then saturates at  0.035 eV which is close to the room temperature. The decrease in the dust kinetic temperature at increasing pressure can be attributed to the collisional cooling of the dust particles by the increasing number of neutral particles. 
%The particle-cloud is illuminated by a horizontally expanded thin sheet of green laser light (532 nm, 100 mW) which is sufficiently constricted vertically to study an individual layer of the dust cloud. The Mie-scattered light from the dust particles is captured by a CCD camera (shown in Fig~\ref{fig:fig1}) at 25 fps with a resolution of 9 $\mu$m/pixel and the images are stored into a high - speed computer.

%\subsection{Comparison with PIV analysis}

%=========================================================================================

\section{Benchmarking in three dimensional simulation\label{3D}}

In order to further test the correctness of the simulation in three dimensions, we carry out extensive numerical comparison between three dimensional Open-MP parallel molecular dynaics and Open-MP parallel Monte-Carlo simulation. To test our numerical simulation we reproduce the Pair Correlation Function, $g(r)$ with radius, $r$, at $\Gamma = 40$ and $120$ reproducing the earlier two results obtained by Farouki and Hamaguchi in 1994 \cite{farouki:1994} from both three dimensional Molecular Dynamics and three dimensional Monte-Carlo simulation.\\

The Pair Correlation Function, $g(r)$ describes the variation of number of particles as a function of radial distance from a reference particle. Let us consider a spherical shell of inner radius $r$ and outer radius $r+\Delta r$. If we count the number of particles within the shell, in principle it gives the pair correlation function of the system. If $h(r)$ be the number of particles within a shell of thickness $dr$ at a distance $r$ then,
\begin{eqnarray*}
g(r) = \frac{L^3 h(r)}{2 \pi r^2 (\Delta r)^3 N^2 }
\end{eqnarray*}
As the system gets more and more ordered, the probability of finding particles at a specific radial distance increases. Hence the oscillations in the $g(r)$ become more prominant and persistent. Thus $g(r)$ is a suitable diagnostics for the identification of phase transition since, during phase transition the system goes from a disordered state to an ordered state.

\begin{figure}[h!]
\includegraphics[scale=0.65]{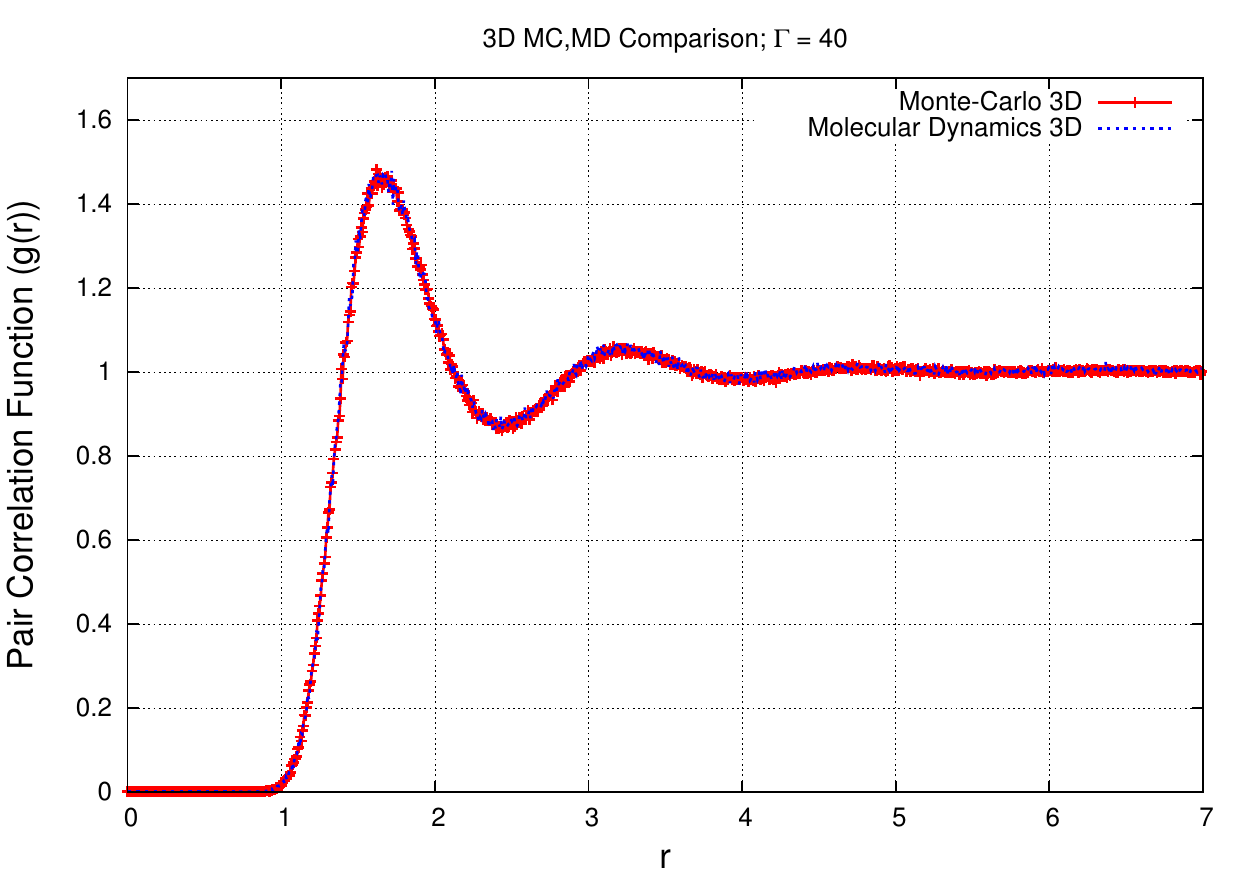}
\caption{Comparison between Pair Correlation Function (g(r)) with radius (r) derived from three dimensional Molecular Dynamics and Monte-Carlo simulation at $\Gamma = 40$ at the micro-canonical run at $N = 5000$, $\kappa = 1$ with periodic boundary condition. The result is found to reproduce the result earlier obtained by Farouki and Hamaguchi \cite{farouki:1994} [Fig. 6].}
\label{MCMD3_40}
\end{figure}
\begin{figure}[h!]
\includegraphics[scale=0.65]{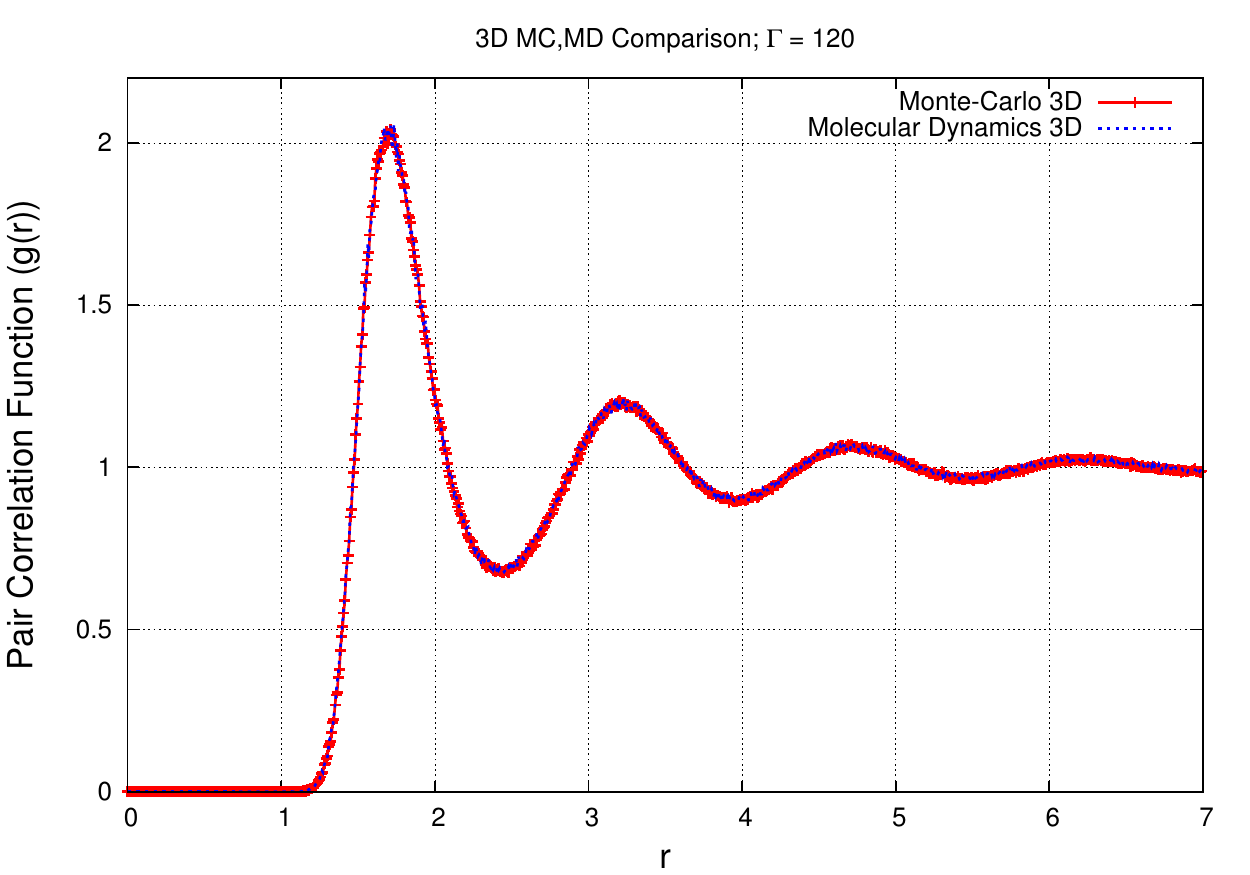}
\caption{Comparison between Pair Correlation Function (g(r)) with radius (r) derived from three dimensional Molecular Dynamics and Monte-Carlo simulation at $\Gamma = 120$ at the micro-canonical run at $N = 5000$, $\kappa = 1$ with periodic boundary condition. The result is found to reproduce the result earlier obtained by Farouki and Hamaguchi \cite{farouki:1994} [Fig. 6].}
\label{MCMD3_120}
\end{figure}

From the extensive numerical simulation it is found that the new diagnostics, `Configurational Temperature' works well in three dimensions as well.

\begin{figure}[h!]
\includegraphics[scale=0.65]{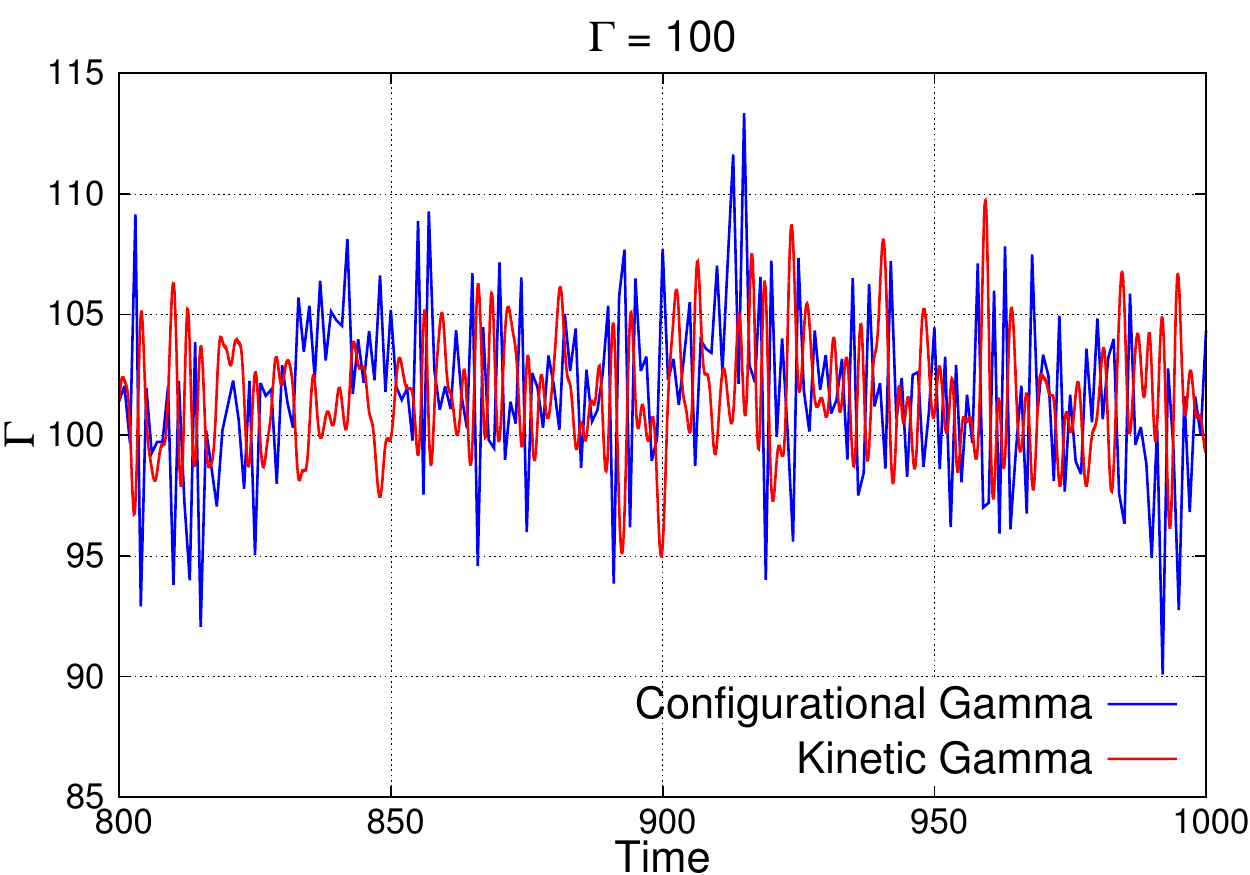}
\caption{Comparison between Configurational $\Gamma$ and Kinetic $\Gamma$ at the micro-canonical run in three dimensional Molecular Dynamics simulation at $N = 5000$, $\kappa = 1$ and $\Gamma = 100$ with periodic boundary condition.}
\label{MD3}
\end{figure}
\begin{figure}[h!]
\includegraphics[scale=0.65]{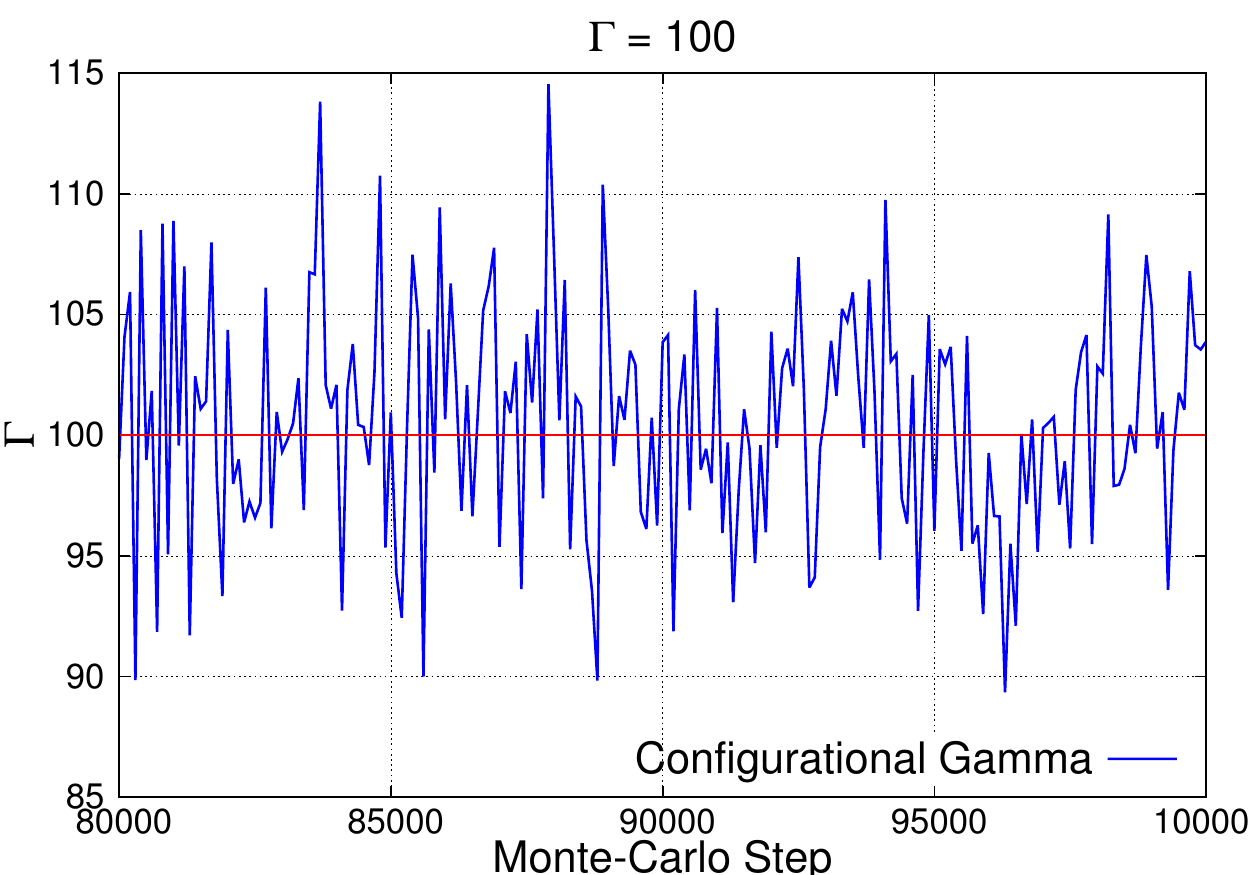}
\caption{Monte-Carlo run with $N = 5000$, $\kappa = 1$, $\Gamma = 100$ and periodic boundary condition.}
\label{MC3}
\end{figure}

%=========================================================================================

\section{Summary and Conclusion\label{conclusion}}

In this paper we report a new diagnostics called `Configurational Temperature' which is found to efficiently reproduce the temperature of a dusty plasma when only the position and interaction potential of the particles are known. The diagnostics is found to work well both in solid and fluid regime in two dimensional Molecular Dynamics simulation and in two dimensional Monte-Carlo simulation. Also it is found that both in periodic as well as in reflecting boundary condition the diagnostics gives good agreement with the expected result for few parameters. However, though in periodic boundary condition at higher coupling strength (lower $\kappa$), the kinetic temperature matches well with `Configurational Temperature', the scenario changes for reflecting boundary condition. We believe that, this is because the interaction potential sees a strong truncation at the boundary. Including the boundary in the interaction potential as a soft-delta-function is an interesting problem and will be reported elsewhere. In three dimensions also, `Configurational Temperature' is found to reproduce the expected temperature from both the Molecular Dynamics and Monte-Carlo simulation. However, in three dimensions, considerable deviation is noted between the kinetic temperature and the `Configurational Temperature' from Molecular Dynamics simulation with reflecting boundary condition. Those results are not presented in this paper. Finally we found that in PIV data analysis from the experimental results from DPEx data, the `Configurational Temperature' shows good agreement at higher pressure. However, in the lower pressure (higher temperature) there are considerable deviations. We believe the model interaction potential at lower pressure may not be modelled well and might need modification. Or the particle tracking at low pressure itself may introduce higher error into the experimental data analysis which keeps us refrain from comparing the two results in detail.\\

Thus in spite of the incapabilities, it is inferred that, the diagnostics will be helpful for measuring dust temperatures at both the fluid and solid regime especially when the velocity of the dust particles are not known. One of such examples are Monte-Carlo simulation where, the simulation data only produces information of the particle position. The difference between particle position in two consecutive Monte Carlo Time frames does not provide any velocity information since the particles do not obey any dynamical equation in this algorithm, rather, are displaced randomly using a standard random number generator with some energy weighted probability ascribed to the individual particle. Hence this novel technique is expected to be a very helpful diagnostics for the community.

%=========================================================================================

\section{Acknowledgement}

All work has been performed in Uday and Udbhav cluster of Institute for Plasma Research, India. RM acknowledges Akanksha Gupta, Indian Institute of Technology, Kanpur for several helpful discussions regarding the implementation of `Configurational Temperature' as a diagnostics in Molecular Dynamics simulation. RM also thanks Harish Charan, Weizmann Institute of Science, for several helpful discussion regarding Molecular Dynamics simulation. S. Jaiswal acknowledge the Institute for plasma research for providing the infrastructure for the experiment and additional funding support provided by the NSF EPSCoR program (OIA-1655280). This work has been supported by the MGK SciDAC projects and DOE Contract DE-AC02-09CH11466.

%=========================================================================================

\bibliography{biblio.bib}
%\begin{thebibliography}{2}
%\bibitem{jaiswal:2019}
%S. Jaiswal and E. Thomas, Plasma Research Express, (2019) https://doi.org/10.1088/2516-1067/ab1f30
%\end{thebibliography}
\end{document}